\documentclass[aps,pra,preprint,superscriptaddress,longbibliography]{revtex4-2}

\usepackage{graphicx}
\usepackage{amsmath}
\usepackage{txfonts}
\usepackage{amsfonts}
\usepackage{amssymb}
\usepackage{color}
\usepackage{color}
\usepackage[pdftex]{hyperref}
\usepackage[normalem]{ulem}

\begin{document}

	\title{Mapping continuous-variable quantum states onto optical scalar beams}
	
	\author{B.~M. Rodr\'iguez-Lara} 
	\email[e-mail: ]{blas.rodriguez@gmail.com}
	
	\author{F.~E. Becerra} 
	\email[email: ]{fbecerra@unm.edu}
	\affiliation{The University of New Mexico, Albuquerque, NM 87106, USA}

	\begin{abstract}
		Optical fields provide an accessible platform to explore connections between classical and quantum mechanics. 
		We introduce a group-theoretic framework based on the $\mathrm{su}(1,1)$ Lie algebra to construct classical analogs of continuous-variable quantum states using the spatial degree of freedom of paraxial scalar beams. 
		Our framework maps squeezed number states onto scalar beams expanded in orthonormal Gaussian modal bases, encompassing both Gaussian and non-Gaussian classical analogs, including one- and two-mode squeezed beams.
		To characterize the structural changes induced by squeezing, we examine phase-space redistribution through Fourier analysis and optical Wigner distribution functions. 
		We derive analytical expressions for the waist, curvature, and Gouy phase of two-mode squeezed Laguerre-Gaussian beams, and establish a relation between the number of accessible modes and the achievable squeezing under finite numerical aperture. 
		While squeezing introduces spatial and spectral correlations that reshape the beam structure, these beams remain constrained by the diffraction limit, as confirmed by the numerical propagation of apodized beams.
		These correlations give rise to classical entanglement. 
		We establish a classical analog of the Duan–Simon inseparability criterion for continuous-variable two-mode Gaussian states. 
		For non-Gaussian squeezed states, we analyze the marginal optical Wigner distribution functions and identify phase-space features, such as negativity, that act as witnesses of classical continuous-variable entanglement.
		Our framework unifies classical analogs of continuous-variable quantum states through beam engineering, enabling quantum-inspired applications in optical imaging, metrology, and communication.
	\end{abstract}
	
	\maketitle
	\newpage
	
	\section{Introduction}
	
	Classical systems can exhibit local correlations between different degrees of freedom that resemble quantum entanglement, a phenomenon often referred to as classical entanglement. 
	In optical fields, such correlations arise when polarization, spatial modes, or orbital angular momentum become intrinsically linked, producing measurement outcomes that mimic quantum behavior \cite{Spreeuw1998, Qian2011, Kagalwala2013}. 
	For instance, single-photon fields and structured light beams display classical entanglement when the coupling between spatial and polarization modes generates non-separable superposition states with intermodal correlations \cite{Toppel2014, Borges2010, Zhao2020}. 
	Although these correlations reproduce statistical signatures of quantum entanglement, they remain local and do not satisfy the non-locality criterion fundamental to quantum mechanics, as demonstrated in Bell-type inequality tests \cite{Bell1964, Zeilinger1999}.
	
	The connection between classical and quantum mechanics—and between classical and quantum correlations, such as classical entanglement—has been studied extensively, with most efforts focused on discrete-variable (DV) systems. 
	In these systems, discrete degrees of freedom (DoFs), such as photon number, polarization, or path encoding, display entanglement-like correlations \cite{Shen2022}. 
	Beyond DV systems, many physical platforms exhibit correlations in continuous-variable (CV) DoFs, including field quadratures, amplitude and phase fluctuations, and Stokes polarization components \cite{Wang2024a, Wang2024b, MoralesRodriguez2024, TorresLeal2024, AguirreOlivas2025}. 
	Understanding how these correlations arise and connect to their quantum counterparts reveals structural features of optical fields and enables quantum-inspired approaches to optical metrology, imaging, and communication.
	
	Recent studies in CV optical systems have explored classical analogs of coherent and squeezed states in structured light. 
	Optical analogs of both coherent and generalized coherent states have been constructed through superpositions of Laguerre-Gaussian (LGB) \cite{MoralesRodriguez2024p1498} and Hermite-Gaussian (HGB) beams \cite{MoralesRodriguez2024}, corresponding to canonical and group-theoretic constructions, respectively. 
	These analogs exploit the symmetries of the two-dimensional paraxial wave equation through Lie group methods, producing statistical features that mirror quantum multiparticle systems \cite{MoralesRodriguez2024p1498, Collado2024, MoralesRodriguez2024, TorresLeal2024, AguirreOlivas2025}. 
	Other approaches use modal superpositions \cite{Shen2022} and geometric beams \cite{Wang2024a, Wang2024b} to demonstrate quadrature-like fluctuations and spatially induced squeezing beyond the standard spatial limit, the classical counterpart to the standard quantum limit. 
	Despite these specific examples, no unified group-theoretic framework connects classical optics with CV quantum states. 
	Establishing such a framework would clarify the structure of classical analogs of quantum correlations in CV DoFs and extend their use to optical imaging, metrology, and communication.
	
	We introduce a group-theoretic framework to construct classical analogs of CV quantum states by exploiting the symmetry structure of the two-dimensional paraxial wave equation. 
	Our approach unifies and generalizes previous methods by mapping the Lie algebra structure of optical fields onto CV quantum properties. 
	It defines a squeezed orthonormal basis for any squeezing parameter, enabling the construction of squeezed versions of arbitrary scalar fields, including single- and two-mode squeezed number states. 
	We apply this formalism to optical imaging and show that diffraction imposes a fundamental limit on the degree of achievable squeezing. 
	We also establish a classical analog of the Duan–Simon quantum inseparability criterion for two-mode optical fields and use the optical Wigner distribution function to identify classical CV entanglement in Gaussian and non-Gaussian states. 
	Our framework provides a complete algebraic description of scalar-field analogs of CV quantum states in paraxial optics and supports quantum-inspired applications in imaging, metrology, and communication.
	
	Our paper is organized as follows. 
	In Sec.~\ref{sec:S2}, we introduce the Lie group approach to squeezing and establish its connection to the special unitary group $\mathrm{SU}(1,1)$, including its representations for single- and two-mode squeezing. This formalism provides the foundation for constructing optical analogs in different representations. 
	We construct explicit examples of squeezed scalar beams in Sec.~\ref{sec:S3}, using Hermite-Gaussian and Laguerre-Gaussian modes. 
	In Sec.~\ref{sec:S4}, we analyze three central consequences of our framework. 
	We examine phase-space redistribution in Sec.~\ref{PhaseRedist} using Fourier analysis and optical Wigner distribution functions. 
	In Sec.~\ref{DiffLimit}, we study the diffraction behavior of two-mode squeezed Laguerre-Gaussian beams under a finite numerical aperture and confirm that squeezing does improve resolution for apertures that resolve the standard Gaussian mode.
	We establish a classical analog of the Duan–Simon quantum inseparability criterion and identify classical CV entanglement in Gaussian and non-Gaussian states using the optical Wigner distribution function in Sec.~\ref{ClassEntanglement}. 
	In Sec.~\ref{Conclusions}, we present our concluding remarks.
	
	\section{SU(1,1) Group and Squeezing}
	\label{sec:S2}
	
	The special unitary group $\mathrm{SU}(1,1)$ plays a central role in physics, from relativity \cite{Carmeli2000} to quantum optics \cite{Wodkiewicz1985,Brif1996}. 
	It is locally isomorphic to the special orthogonal group $\mathrm{SO}(2,1)$, which preserves the quadratic form $x_{1}^{2} + x_{2}^{2} - x_{3}^{2}$ \cite{Bargmann1946, Gilmore2012}. 
	This symmetry leaves invariant a one-parameter family of one-sheeted hyperboloids, $x_{1}^{2} + x_{2}^{2} - x_{3}^{2} = R^{2}$, the light cone, $x_{1}^{2} + x_{2}^{2} - x_{3}^{2} = 0$, and a family of two-sheeted hyperboloids, $x_{1}^{2} + x_{2}^{2} - x_{3}^{2} = -R^{2}$. 
	These surfaces extend indefinitely, reflecting the non-compact character of $\mathrm{SU}(1,1)$. 
	The invariance of the light cone under $\mathrm{SU}(1,1)$ transformations connects the group directly to the Lorentz group in $(2+1)$ dimensions \cite{Naber2012}. 
	In quantum optics, the $\mathrm{SU}(1,1)$ group provides the algebraic structure for squeezed states \cite{Drummond2004}, which support applications in quantum sensing \cite{Degen2017}, metrology \cite{Pezze2018, Gessner2019}, and information processing \cite{Weedbrook2012, Toth2014}.
	
	Elements of the $\mathrm{SU}(1,1)$ group are expressed as exponential operators,
	\begin{align}
		\hat{S}(\xi) = e^{i \zeta_{j} \hat{K}_{j}},
	\end{align}
	where $\zeta_{j}$ is a real parameter and $\hat{K}_{j}$ are the generators of the group. 
	These generators span the tangent space near the identity, satisfy the commutation relations \cite{Gilmore2012},
	\begin{align}
		\begin{aligned}
			\left[ \hat{K}_{0}, \hat{K}_{1} \right] &= i \hat{K}_{2}, \\
			\left[ \hat{K}_{1}, \hat{K}_{2} \right] &= -i \hat{K}_{0}, \\
			\left[ \hat{K}_{2}, \hat{K}_{0} \right] &= i \hat{K}_{1},
		\end{aligned}
		\label{s11algebra}
	\end{align}
	and define the $\mathrm{su}(1,1)$ Lie algebra.
	
	The generator $\hat{K}_{0}$ induces cyclic transformations in the $(\hat{K}_{1}, \hat{K}_{2})$ plane,
	\begin{align} 
		\begin{aligned} 
			e^{i \zeta_{0} \hat{K}_{0}} \hat{K}_{1} e^{-i \zeta_{0} \hat{K}_{0}} &= \hat{K}_{1} \cos \zeta_{0} - \hat{K}_{2} \sin \zeta_{0}, \\
			e^{i \zeta_{0} \hat{K}_{0}} \hat{K}_{2} e^{-i \zeta_{0} \hat{K}_{0}} &= \hat{K}_{2} \cos \zeta_{0} + \hat{K}_{1} \sin \zeta_{0}, 
		\end{aligned} 
	\end{align}
	with period $2\pi$ for $\zeta_{0} \in [0, 2\pi)$. In contrast, the generators $\hat{K}_{1}$ and $\hat{K}_{2}$ induce hyperbolic transformations,
	\begin{align} 
		\begin{aligned} 
			e^{i \zeta_{1} \hat{K}_{1}} \hat{K}_{0} e^{-i \zeta_{1} \hat{K}_{1}} &= \hat{K}_{0} \cosh \zeta_{1} + \hat{K}_{2} \sinh \zeta_{1}, \\
			e^{i \zeta_{1} \hat{K}_{1}} \hat{K}_{2} e^{-i \zeta_{1} \hat{K}_{1}} &= \hat{K}_{2} \cosh \zeta_{1} + \hat{K}_{0} \sinh \zeta_{1}, \\
			e^{i \zeta_{2} \hat{K}_{2}} \hat{K}_{0} e^{-i \zeta_{2} \hat{K}_{2}} &= \hat{K}_{0} \cosh \zeta_{2} - \hat{K}_{1} \sinh \zeta_{2}, \\
			e^{i \zeta_{2} \hat{K}_{2}} \hat{K}_{1} e^{-i \zeta_{2} \hat{K}_{2}} &= \hat{K}_{1} \cosh \zeta_{2} - \hat{K}_{0} \sinh \zeta_{2},
		\end{aligned} 
	\end{align}
	corresponding to hyperbolic translations in the $(\hat{K}_{0}, \hat{K}_{1})$ and $(\hat{K}_{0}, \hat{K}_{2})$ planes.
	The unbounded nature of the parameters $\zeta_{1}, \zeta_{2} \in [0, \infty)$ reflects the non-compact character of the group. 
	This non-compactness is also evident in the spectra of the generators, where $\hat{K}_{0}$ has a discrete spectrum while $\hat{K}_{1}$ and $\hat{K}_{2}$ have continuous spectra \cite{Puri2012}.
	
	The Casimir operator of the $\mathrm{su}(1,1)$ algebra,
	\begin{align} 
		\hat{K}^{2} = \hat{K}_{0}^{2} - \hat{K}_{1}^{2} - \hat{K}_{2}^{2}, 
	\end{align} 
	commutes with all generators,
	\begin{align} 
		\left[ \hat{K}^{2}, \hat{K}_{j} \right] = 0,  \qquad j = 0,1,2, 
	\end{align}
	and remains invariant under group transformations. This property defines a class of representations where both $\hat{K}^{2}$ and the discrete-spectrum generator $\hat{K}_{0}$ are diagonal \cite{Puri2012},
	\begin{align} 
		\begin{aligned} 
			\hat{K}^{2} \, \vert k, n \rangle &= k (k - 1) \, \vert k, n \rangle, \\
			\hat{K}_{0} \, \vert k, n \rangle &= (k + n) \, \vert k, n \rangle,
		\end{aligned} 
	\end{align}
	where $k > 0$ is the Bargmann index and $n = 0, 1, 2, \ldots$ labels the states in the orthonormal basis of the representation.
	
	The ladder operators,
	\begin{align}
		\hat{K}_{\pm} = \hat{K}_{1} \pm i \hat{K}_{2},
	\end{align}
	raise and lower the Bargmann states,
	\begin{align}
		\begin{aligned}
			\hat{K}_{+} \, \vert k, n \rangle &= \sqrt{(n+1)(2k + n)} \, \vert k, n + 1 \rangle, \\
			\hat{K}_{-} \, \vert k, n \rangle &= \sqrt{n (2k + n - 1)} \, \vert k, n - 1 \rangle,
		\end{aligned}
	\end{align}
	in that order.
	The $\mathrm{su}(1,1)$ commutation relations in terms of the Cartan and ladder operators,
	\begin{align} 
		\begin{aligned} 
			\left[ \hat{K}_{0}, \hat{K}_{\pm} \right] &= \pm \hat{K}_{\pm}, \\ 
			\left[ \hat{K}_{+}, \hat{K}_{-} \right] &= -2 \hat{K}_{0}, 
		\end{aligned} 
	\end{align}
	mirror the structure of the $\mathrm{su}(2)$ Lie algebra,
	\begin{align} 
		\begin{aligned}
			\left[ \hat{J}_{0}, \hat{J}_{\pm} \right] &= \pm \hat{J}_{\pm}, \\ 
			\left[ \hat{J}_{+}, \hat{J}_{-} \right] &= 2 \hat{J}_{0},     
		\end{aligned}
	\end{align}
	with $\mathrm{su}(1,1)$ serving as the non-compact analog, where hyperbolic transformations replace compact rotations.

	\subsection{Single-mode squeezing}
	
	An explicit connection between the abstract $\mathrm{su}(1,1)$ Lie algebra and physical systems emerges in the one-dimensional harmonic oscillator,
	\begin{align}
		\frac{1}{2} \left( \hat{p}_{x}^{2} + \hat{q}_{x}^{2} \right) \psi_{n}(q_{x}) = \left(n + \frac{1}{2} \right) \psi_{n}(q_{x}),
	\end{align}
	where $\hat{p}_{x} = -i \partial_{q_{x}}$ is the dimensionless momentum operator, $\hat{q}_{x}$ is the dimensionless position operator, and $n = 0, 1, 2, \ldots$ labels the energy eigenstates. 
	The eigenfunctions,
	\begin{align}
		\psi_{n}(q_{x}) = \sqrt{\frac{1}{\sqrt{\pi} 2^{n} n!}} \, e^{-\frac{1}{2} q_{x}^{2}} H_{n}(q_{x}),
	\end{align}
	are Hermite-Gauss modes with Hermite polynomials $H_{n}(q_{x})$ \cite{Griffiths2018}.
	
	The dimensionless canonical commutator,
	\begin{align}
		\left[ \hat{q}_{x}, \hat{p}_{x} \right] = i,
	\end{align}
	defines the symplectic algebra $\mathrm{sp}(2, \mathbb{R})$, which contains two realizations of the $\mathrm{su}(1,1)$ algebra,
	\begin{align}
		\begin{aligned}
			\hat{K}_{0} &= \frac{1}{4} \left( \hat{p}_{x}^{2} + \hat{q}_{x}^{2} \right), \\
			\hat{K}_{1} &= \frac{1}{4} \left( \hat{p}_{x}^{2} - \hat{q}_{x}^{2} \right), \\
			\hat{K}_{2} &= \frac{1}{4} \left\{ \hat{q}_{x}, \hat{p}_{x} \right\},
		\end{aligned}
	\end{align}
	where $\left\{ \hat{q}_{x}, \hat{p}_{x} \right\} = \hat{q}_{x} \hat{p}_{x} + \hat{p}_{x} \hat{q}_{x}$ is the anticommutator. 
	These generators satisfy the $\mathrm{su}(1,1)$ commutation relations.
	The operator $\hat{K}_{0}$ corresponds to the harmonic oscillator, while $\hat{K}_{1}$ corresponds to the inverted oscillator,
	\begin{align}
		2 \hat{K}_{1} \, \phi_{\lambda}(q_{x}) = \left( \hat{p}_{x}^{2} - \hat{q}_{x}^{2} \right) \phi_{\lambda}(q_{x}) = \lambda \, \phi_{\lambda}(q_{x}),
	\end{align}
	with parabolic cylinder eigenfunctions and continuous spectrum $\lambda \in \mathbb{R}$. 
	The continuous spectrum of $\hat{K}_{2}$ follows from that of $\hat{K}_{1}$, since a $\pi/2$ rotation generated by $\hat{K}_{0}$ in the $(\hat{K}_{1}, \hat{K}_{2})$ plane transforms $\hat{K}_{1}$ into $\hat{K}_{2}$.
	The combination $\hat{K}_{0} + \hat{K}_{1}$ defines the eigenvalue problem,
	\begin{align}
		2 (\hat{K}_{0} + \hat{K}_{1}) \, \phi_{\lambda}(q_{x}) = -\partial_{q_{x}}^{2} \phi_{\lambda}(q_{x}) = \lambda \, \phi_{\lambda}(q_{x}),
	\end{align}
	with plane wave eigenfunctions and continuous spectrum $\lambda \in \mathbb{R}$. 
	
	The Casimir operator for this representation,
	\begin{align}
		\hat{K}^{2} = \hat{K}_{0}^{2} - \hat{K}_{1}^{2} - \hat{K}_{2}^{2} = -\frac{3}{16},
	\end{align}
	implies two possible Bargmann indices, $k = 1/4$ and $k = 3/4$, which both satisfy $k(k - 1) = -3/16$. 
	These values label two irreducible representations,
	\begin{align}
		\begin{aligned}
			\langle q_{x} \vert k = 1/4, n \rangle &= \psi_{2n}(q_{x}), \\
			\langle q_{x} \vert k = 3/4, n \rangle &= \psi_{2n+1}(q_{x}),
		\end{aligned}
	\end{align}
	associated with even and odd parity, respectively. 
	The $\mathrm{su}(1,1)$ symmetry thus partitions the harmonic oscillator states into two parity-separated ladders.
	
	Squeezing introduces a correlated rescaling of position and momentum. 
	The group element generated by $\hat{K}_{2}$ transforms the canonical pair,
	\begin{align}
		\begin{aligned}
			e^{i \zeta_{2} \hat{K}_{2}} \, \hat{q}_{x} \, e^{-i \zeta_{2} \hat{K}_{2}} &= \hat{q}_{x} e^{\frac{1}{2} \zeta_{2}}, \\
			e^{i \zeta_{2} \hat{K}_{2}} \, \hat{p}_{x} \, e^{-i \zeta_{2} \hat{K}_{2}} &= \hat{p}_{x} e^{-\frac{1}{2} \zeta_{2}},
		\end{aligned}
	\end{align}
	expanding one quadrature while compressing the other. 
	The squeezed eigenstates of the one-dimensional harmonic oscillator,
	\begin{align}
		\psi_{n,\zeta_{2}}(q_{x}) = e^{-i \zeta_{2} \hat{K}_{2}} \psi_{n}(q_{x}),
	\end{align}
	have variances
	\begin{align}
		\begin{aligned}
			\langle (\Delta q_{x})^{2} \rangle &= \left( n + \frac{1}{2} \right) e^{\zeta_{2}}, \\
			\langle (\Delta p_{x})^{2} \rangle &= \left( n + \frac{1}{2} \right) e^{-\zeta_{2}},
		\end{aligned}
		\label{SqVariances}
	\end{align}
	showing increased uncertainty in one quadrature and reduced uncertainty in the conjugate while preserving the Heisenberg bound, which only the squeezed vacuum state, $n = 0$, saturates.

	\subsection{Two-mode squeezing}
	
	An explicit connection between the $\mathrm{su}(1,1)$ Lie algebra and two-mode systems emerges in the two-dimensional quantum harmonic oscillator in Cartesian coordinates,
	\begin{align}
		\frac{1}{2} \left( \hat{p}_{x}^{2} + \hat{q}_{x}^{2} + \hat{p}_{y}^{2} + \hat{q}_{y}^{2} \right) \psi_{n_{x},n_{y}}(q_{x}, q_{y}) = \left( n_{x} + n_{y} + 1 \right) \psi_{n_{x},n_{y}}(q_{x}, q_{y}),
	\end{align}
	 where $\hat{q}_{j}$ and $\hat{p}_{j} = -i \partial_{q_{j}}$ are dimensionless position and momentum operators for $j = x, y$. 
	The eigenfunctions,
	\begin{align}
		\psi_{n_{x},n_{y}}(q_{x},q_{y}) = \frac{1}{\sqrt{\pi 2^{n_{x}+n_{y}} n_{x}! n_{y}!}} e^{- \frac{1}{2} (q_{x}^{2} + q_{y}^{2})} H_{n_{x}}(q_{x}) H_{n_{y}}(q_{y}),
	\end{align}
	are products of Hermite-Gauss modes that follow from the separability of the Hamiltonian in Cartesian coordinates.
	
	An equivalent expression for the eigenfunctions arises in polar coordinates for systems with cylindrical symmetry. 
	The two-dimensional harmonic oscillator becomes,
	\begin{align}
		\frac{1}{2} \left( \hat{p}_{x}^{2} + \hat{q}_{x}^{2} + \hat{p}_{y}^{2} + \hat{q}_{y}^{2} \right) \psi_{p,\ell}(q_{\rho}, q_{\theta}) = \left( 2p + \vert \ell \vert + 1 \right) \psi_{p, \ell}(q_{\rho}, q_{\theta}),
	\end{align}
	where the eigenfunctions,
	\begin{align}
		\psi_{p,\ell}(q_{\rho}, q_{\theta}) = (-1)^{p} \sqrt{\frac{p!}{\pi (p + \vert \ell \vert)!}} \, q_{\rho}^{\vert \ell \vert} e^{- \frac{1}{2} q_{\rho}^{2}} \mathrm{L}_{p}^{\vert \ell \vert} \left(q_{\rho}^{2}\right) e^{i \ell q_{\theta}},
	\end{align}
	are Laguerre-Gauss modes expressed in terms of the radial and azimuthal quantum numbers \(p\) and \(\ell\), and the generalized Laguerre polynomials \(\mathrm{L}_{p}^{\vert \ell \vert}(\cdot)\) \cite{CohenTannoudji1977}.
	
	The dimensionless canonical commutation relations,
	\begin{align}
		\left[ \hat{q}_{j}, \hat{p}_{j} \right] = i, \qquad j = x, y,
	\end{align}
	define a two-mode realization of the symplectic algebra $\mathrm{sp}(4, \mathbb{R})$, which contains an embedded realization of the $\mathrm{su}(1,1)$ algebra,
	\begin{align}
		\begin{aligned}
			\hat{K}_{0} &= \frac{1}{4} \left( \hat{p}_{x}^{2} + \hat{q}_{x}^{2} + \hat{p}_{y}^{2} + \hat{q}_{y}^{2} \right), \\
			\hat{K}_{1} &= \frac{1}{2} \left( \hat{q}_{x} \hat{q}_{y} - \hat{p}_{x} \hat{p}_{y} \right), \\
			\hat{K}_{2} &= -\frac{1}{2} \left( \hat{q}_{x} \hat{p}_{y} + \hat{q}_{y} \hat{p}_{x} \right).
		\end{aligned}
	\end{align}
	Here, $\hat{K}_{0}$ corresponds to the total energy of the uncoupled oscillators, and $\hat{K}_{1}$ and $\hat{K}_{2}$ generate two-mode squeezing transformations.
	
	The Casimir operator for this representation,
	\begin{align}
		\hat{K}^{2} = \frac{1}{4} \left[ \left( \hat{p}_{x}^{2} + \hat{q}_{x}^{2} - \hat{p}_{y}^{2} - \hat{q}_{y}^{2} \right)^{2} - 1 \right],
	\end{align}
	yields Bargmann indices
	\begin{align}
		k = \frac{1}{2} \left( \vert n_{x} - n_{y} \vert + 1 \right),
	\end{align}
	with each pair $(n_{x}, n_{y})$ defining a separate $\mathrm{su}(1,1)$ ladder. 
	For $k = 1/2$, the states take the form
	\begin{align}
		\langle q_{x}, q_{y} \vert k = 1/2, n \rangle = \psi_{n,n}(q_{x}, q_{y}),
	\end{align}
	while all other indices $k = m/2$, with $m = 1, 2, 3, \ldots$, yield the families
	\begin{align}
		\begin{aligned}
			\langle q_{x}, q_{y} \vert k = m/2, n \rangle &= \psi_{n,n+m}(q_{x}, q_{y}), \\
			\langle q_{x}, q_{y} \vert k = m/2, n \rangle &= \psi_{n+m,n}(q_{x}, q_{y}),
		\end{aligned}
	\end{align}
	 which correspond to positive and negative values of $(n_{x} - n_{y})$ and define two orthogonal subspaces for each $k$ \cite{MoralesRodriguez2024}.
	
	Two-mode squeezing transformations arise from the action of the group element generated by $\hat{K}_{2}$ on the canonical operators,
	\begin{align}
		\begin{aligned}
			e^{i \zeta_{2} \hat{K}_{2}} \, \hat{q}_{x} \, e^{-i \zeta_{2} \hat{K}_{2}} &= \hat{q}_{x} \cosh \tfrac{\zeta_{2}}{2} - \hat{q}_{y} \sinh \tfrac{\zeta_{2}}{2}, \\
			e^{i \zeta_{2} \hat{K}_{2}} \, \hat{q}_{y} \, e^{-i \zeta_{2} \hat{K}_{2}} &= \hat{q}_{y} \cosh \tfrac{\zeta_{2}}{2} - \hat{q}_{x} \sinh \tfrac{\zeta_{2}}{2}, \\
			e^{i \zeta_{2} \hat{K}_{2}} \, \hat{p}_{x} \, e^{-i \zeta_{2} \hat{K}_{2}} &= \hat{p}_{x} \cosh \tfrac{\zeta_{2}}{2} + \hat{p}_{y} \sinh \tfrac{\zeta_{2}}{2}, \\
			e^{i \zeta_{2} \hat{K}_{2}} \, \hat{p}_{y} \, e^{-i \zeta_{2} \hat{K}_{2}} &= \hat{p}_{y} \cosh \tfrac{\zeta_{2}}{2} + \hat{p}_{x} \sinh \tfrac{\zeta_{2}}{2},
		\end{aligned}
	\end{align}
	mixing the canonical variables and correlating the quadratures of the two oscillators.
	
	\section{Squeezed Paraxial Scalar Beams}
	\label{sec:S3}
	We began in Sec.~\ref{sec:S2} with explicit representations of the $\mathrm{su}(1,1)$ algebra using one- and two-dimensional quantum harmonic oscillators. 
	These examples, standard in quantum optics, provide a concrete setting for squeezing in Cartesian and polar coordinates. 
	However, the algebraic structure of $\mathrm{su}(1,1)$ extends beyond these particular cases and supports abstract realizations, independent of the coordinate system.
	
	We now introduce an abstract group-theoretic framework based on the $\mathrm{su}(1,1)$ algebra. 
	In the single-mode case, the generators are
	\begin{align}
		\begin{aligned}
			\hat{K}_{0} &= \frac{1}{4} \left( \hat{p}_{1}^{2} + \hat{q}_{1}^{2} \right), \\
			\hat{K}_{1} &= \frac{1}{4} \left( \hat{p}_{1}^{2} - \hat{q}_{1}^{2} \right), \\
			\hat{K}_{2} &= \frac{1}{4} \left\{ \hat{q}_{1}, \hat{p}_{1} \right\},
		\end{aligned}
	\end{align}
	while the two-mode case uses the generators
	\begin{align}
		\begin{aligned}
			\hat{K}_{0} &= \frac{1}{4} \left( \hat{p}_{1}^{2} + \hat{q}_{1}^{2} + \hat{p}_{2}^{2} + \hat{q}_{2}^{2} \right), \\
			\hat{K}_{1} &= \frac{1}{2} \left( \hat{q}_{1} \hat{q}_{2} - \hat{p}_{1} \hat{p}_{2} \right), \\
			\hat{K}_{2} &= -\frac{1}{2} \left( \hat{q}_{1} \hat{p}_{2} + \hat{q}_{2} \hat{p}_{1} \right),
		\end{aligned}
	\end{align}
	in terms of dimensionless canonical pairs with commutation relations $[\hat{q}_j, \hat{p}_j] = i$ for $j = 1, 2$.
	
	We use this algebraic structure to define squeezed number states, independently of the representation \cite{Perelomov2012, Puri2012},
	\begin{align}
		\begin{aligned}
			\vert k; \zeta, m \rangle =&~ e^{ \zeta \hat{K}_{+} - \zeta^{\ast}  \hat{K}_{-} } \vert k; m \rangle \\
			=&~ \sum_{j=0}^{\infty} c_{j}(k,\zeta,m) \, \vert k; j \rangle,
		\end{aligned}
	\end{align}
	with expansion coefficients \cite{Marian1991,VillanuevaVergara2015,MoralesRodriguez2024p1498,MoralesRodriguez2024},
	\begin{align}
		\begin{aligned}
			c_{j}(k,\zeta,m) =&~ (-1)^{j} \sqrt{ \frac{\Gamma(2k)}{\Gamma(m+1)\Gamma(2k-m)} \frac{\Gamma(2k)}{\Gamma(j+1)\Gamma(2k-j)} } \times \\
			&~ \times e^{-i \beta (m - j)} \left( \mathrm{sech} \, \alpha \right)^{2k} \left( \mathrm{tanh} \, \alpha \right)^{m+j} \times \\
			&~ \quad \times {}_{2}\mathrm{F}_{1} \left( -j, -m; 2k; -\mathrm{csch}^{2} \alpha \right),
		\end{aligned}
	\end{align}
	in terms of the complex squeezing parameter $\zeta = \alpha e^{i \beta}$, with $\alpha \geq 0$ and $\beta \in [0, 2\pi)$. 
	This formulation supports both single- and two-mode squeezing transformations used in quantum optics \cite{Caves1985, Schumaker1985, Gerry1991}.
	
	To construct squeezed paraxial scalar beams, we start by identifying optical fields that correspond to the Bargmann basis states $\vert k; j \rangle$ in a given representation. 
	In quantum mechanics, this basis spans the Hilbert space and enables a natural decomposition of squeezed states. 
	Similarly, The optical analog requires a complete orthonormal set of paraxial modes that solve the wave equation on each transverse plane. 
	Once this basis is fixed, the squeezed beams follow directly from the group-theoretic expansion.
	
	In practice, researchers often adopt quadrature assignments that simplify the modal structure. 
	Choosing
	\begin{align}
		\begin{aligned}
			\hat{q}_{1} = \hat{q}_{x}, \quad \hat{p}_{1} = \hat{p}_{x}, \quad
			\hat{q}_{2} = \hat{q}_{y}, \quad \hat{p}_{2} = \hat{p}_{y},
		\end{aligned}
	\end{align}
	leads to Hermite-Gauss beams (HGBs), while
	\begin{align}
		\begin{aligned}
			\hat{q}_{1} = \hat{q}_{x} + i \hat{q}_{y}, \quad \hat{p}_{1} = \hat{p}_{x} + i \hat{p}_{y}, \\
			\hat{q}_{2} = \hat{q}_{x} - i \hat{q}_{y}, \quad \hat{p}_{2} = \hat{p}_{x} - i \hat{p}_{y},
		\end{aligned}
	\end{align}
	leads to Laguerre-Gauss beams (LGBs). 
	These assignments preserve the algebraic structure and yield well-known modal bases. 
	The algebraic framework remains valid for any complete orthonormal set that solves the paraxial wave equation; for instance, generalized $\mathrm{su}(2)$ coherent states that interpolate between HGBs and LGBs \cite{AguirreOlivas2025}.
	
	\begin{figure}
		\centering
		\includegraphics[scale = .52]{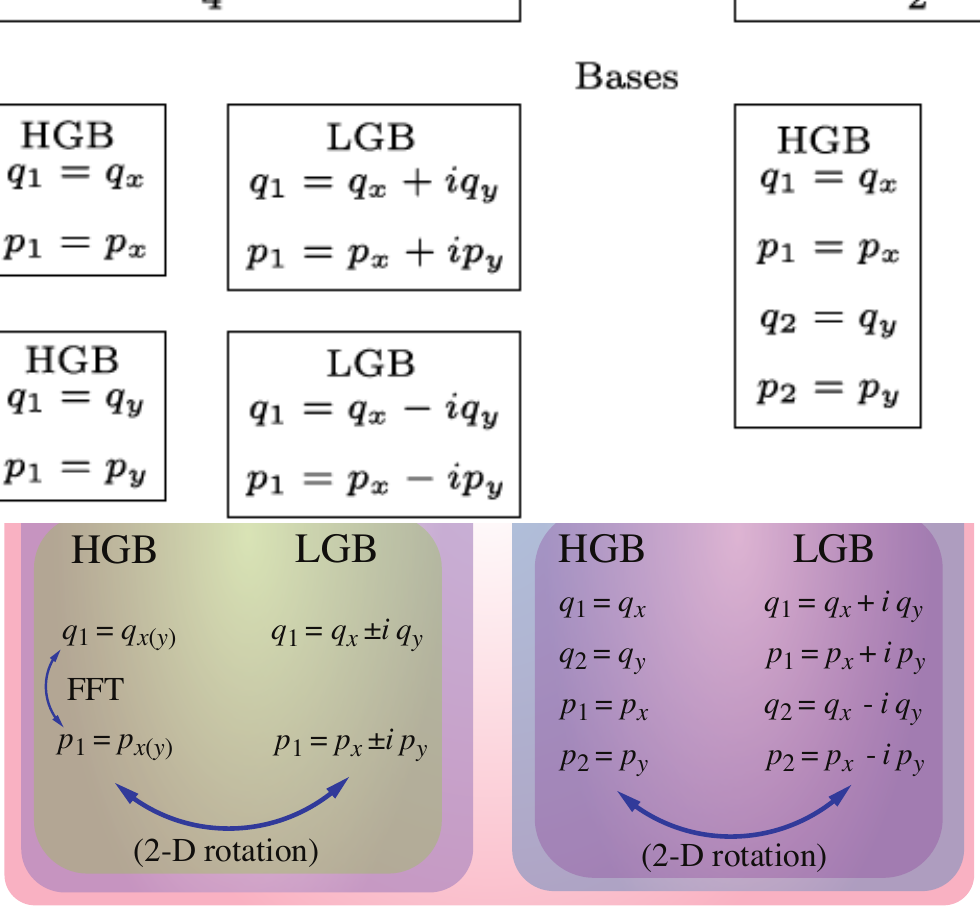}
		\caption{Hierarchy of the framework based on the $\mathrm{su}(1,1)$ algebra to construct squeezed scalar beams.}
		\label{fig:Fig1}
	\end{figure}
	
	Figure~\ref{fig:Fig1} shows the hierarchy of our framework, with abstract single- and two-mode $\mathrm{su}(1,1)$ representations forming the top layer, and a continuum of orthonormal modal bases forming the bottom layer. 
	As an explicit example, we choose generalized $\mathrm{su}(2)$ coherent states that interpolate continuously between HGBs and LGBs~\cite{AguirreOlivas2025}. 
	The choice of an explicit basis determines the spatial structure and correlations of the resulting squeezed beams.
	For clarity and to establish a common reference, we adopt HGBs and LGBs throughout the remainder of this work to provide explicit examples of the abstract formalism.
	
	\subsection{Paraxial Scalar Beams}
	
	As a first step, we define the two standard families of scalar optical beams. 
	The orthonormal HGBs,
	\begin{align}
		\Psi_{n_{x},n_{y}}(x,y,z) = \frac{\sqrt{2}}{w(z)} e^{-\frac{i k (x^{2} + y^{2})}{2 R(z)}} e^{i (n_{x} + n_{y} + 1) \varphi(z)} \psi_{n_{x},n_{y}} \left( \frac{\sqrt{2} x}{w(z)}, \frac{\sqrt{2} y}{w(z)} \right),
	\end{align}
	and LGBs,
	\begin{align}
		\Psi_{p,\ell}(\rho,\theta,z) = \frac{\sqrt{2}}{w(z)} e^{-\frac{i k \rho^{2}}{2 R(z)}} e^{i (2p + \vert \ell \vert + 1) \varphi(z)} \psi_{p,\ell} \left( \frac{\sqrt{2} \rho}{w(z)}, \theta \right),
	\end{align}
	are solutions to the paraxial wave equation and form complete, orthonormal sets across each transverse $z$-plane. 
	The Cartesian modal numbers $n_x$ and $n_y$ determine the number of nodes along the horizontal and vertical directions, while the polar modal numbers $p$ and $\ell$ define the radial and azimuthal node structure, respectively.
	
	We use standard Gaussian beam parameters. 
	The beam waist, radius of curvature, and Gouy phase are
	\begin{align}
		\begin{aligned}
			w(z) =&~ w_0 \sqrt{1 + (z/z_R)^2}, \\
			R(z) =&~ z \left[ 1 + (z_R/z)^2 \right], \\
			\varphi(z) =&~ \tan^{-1}(z / z_R),
		\end{aligned}
	\end{align}
	in that order \cite{Siegman1986}. 
	Here, $w_0$ is the beam waist at $z = 0$, $z_R = \pi w_0^2/\lambda$ is the Rayleigh range, $\lambda$ is the wavelength, and $k = 2\pi/\lambda$ is the wavenumber.
	
	We choose HGBs and LGBs to ease the connection of this work to current literature, as these bases provide a clear and accessible reference for the reader.
	
	\subsection{Single-mode Squeezed Scalar Beams}
	
	We construct optical analogs of single-mode squeezed states~\cite{MoralesRodriguez2024p1498, MoralesRodriguez2024} using coherent superpositions of HGBs,
	\begin{align}
		\begin{aligned}
			\Psi_{n_{x},\zeta_{x},n_{y},\zeta_{y}} (x,y,z) =&~ \sum_{u,v=0}^{\infty} c_{u}(n_{x}, \alpha_{x}, \beta_{x}) c_{v}(n_{y}, \alpha_{y}, \beta_{y}) \times \\
			&~ \times \Psi_{2u + (n_{x} \, \mathrm{mod} \,  2), 2v + (n_{y} \, \mathrm{mod} \,  2)}(x, y, z),
		\end{aligned}
	\end{align}
	where the horizontal and vertical mode numbers, $n_{x}$ and $n_{y}$, and the complex squeezing parameters, $\zeta_{x} = \alpha_{x} e^{i \beta_{x}}$ and $\zeta_{y} = \alpha_{y} e^{i \beta_{y}}$, define independent single-mode squeezing transformations in each direction~\cite{HuertaAlderete2019}. 
	
	The coefficients,
	\begin{align}
		\begin{aligned}
			c_{j}(n, \alpha, \beta) =&~ (-1)^{j} \sqrt{\frac{\Gamma\left[ 2k(n) + n \right]}{\Gamma(n+1)\Gamma\left[ 2k(n) \right]} \frac{\Gamma\left[ 2k(n) + j \right]}{\Gamma(j+1)\Gamma\left[ 2k(n) \right]}} \times \\
			&~ \times e^{i \beta (n - j)} \left( \mathrm{sech} \, \alpha \right)^{2k(n)} \left( \mathrm{tanh} \, \alpha \right)^{n + j} \times \\
			&~ \times {}_{2}F_{1} \left( -j, -n; 2k(n); -\mathrm{csch}^{2} \alpha \right),
		\end{aligned}
	\end{align}
	are the modal weights, with Bargmann parameter $k(n) = 1/4 + (n \, \mathrm{mod} \, 2)/2$, giving $k = 1/4$ for even $n$ and $k = 3/4$ for odd $n$. 
	Here $\Gamma(a)$ is the Gamma function and ${}_{2}F_{1}(a, b; c; d)$ is the Gauss hypergeometric function.
	
	For fixed squeezing parameters, the resulting squeezed HGBs form a new orthonormal basis. 
	As a result, the set of squeezed HGBs for given $\zeta_{x}$ and $\zeta_{y}$ can represent any paraxial beam at that specific value of the squeezing parameter, including optical analogs of general single-mode Gaussian and non-Gaussian CV quantum states.
	
\begin{figure}
		\centering
		\includegraphics[scale = 1]{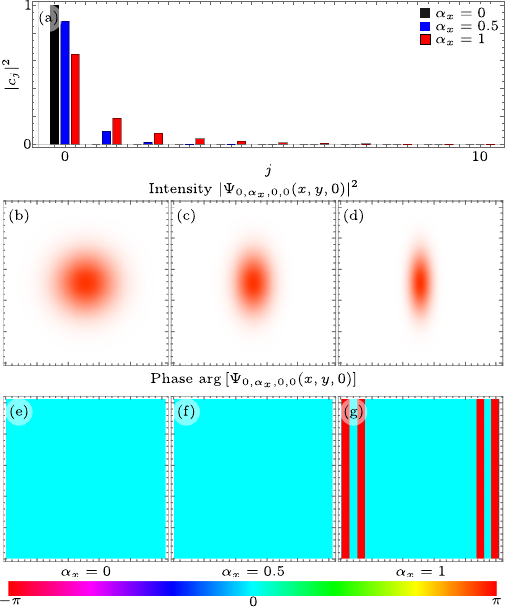}
		\caption{Single-mode squeezed HGB with $(n_{x}, n_{y}) = (0, 0)$. 
			(a) Modal weights $\vert c_{j}(n, \alpha, \beta)\vert^{2}$. (b)--(d) Intensity $\vert \Psi_{n_{x},\zeta_{x},n_{y},\zeta_{y}}(x,y,z) \vert^{2}$ at $z=0$, and (e)--(g) Phase $\mathrm{arg} \left[ \Psi_{n_{x},\zeta_{x},n_{y},\zeta_{y}}(x,y,z) \right]$ at $z=0$, for $\alpha_{x} \in \left\{ 0, 0.5, 1 \right\}$ and $\alpha_{y} = 0$. 
			Beam parameters: $w_{0} = 1~\mathrm{mm}$, $\lambda = 632~\mathrm{nm}$, $\vert x \vert, \vert y \vert \leq 2 w_{0}$.}
		\label{fig:Fig2}
	\end{figure}

\begin{figure}
		\centering
		\includegraphics[scale = 1]{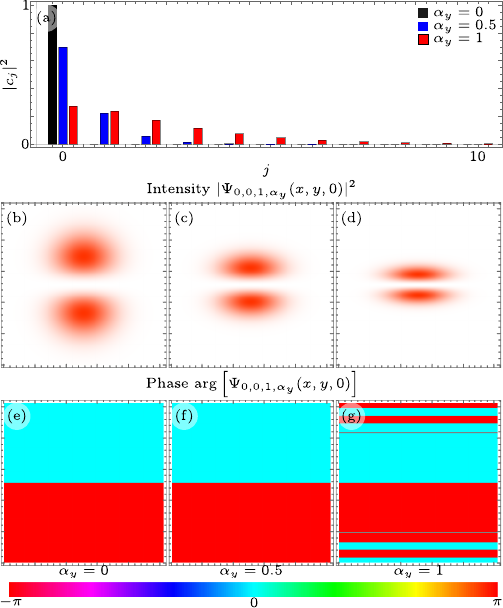}
		\caption{Same as Fig.~\ref{fig:Fig2}, but for a single-mode squeezed HGB with $(n_{x}, n_{y}) = (0, 1)$, $\alpha_{x}=0$, and $\alpha_{y} \in \left\{ 0, 0.5, 1 \right\}$.}
		\label{fig:Fig3}
	\end{figure}

	Figure~\ref{fig:Fig2} shows a single-mode squeezed HGB with $(n_x, n_y) = (0, 0)$, corresponding to the squeezed vacuum state. 
	Figure~\ref{fig:Fig3} shows the case $(n_x, n_y) = (0, 1)$, representing the optical analog of a squeezed single-photon Fock state along $y$. 
	In both figures, panel (a) shows the squared modal weights $\vert c_j \vert^2$ for squeezing strengths $\alpha_x \in \{ 0, 0.5, 1 \}$; panels (b)--(d) show the intensity profiles; and panels (e)--(g) show the phase profiles at $z = 0$ for the same values of $\alpha_x$. 
	Panel (g) highlights the effects of truncating the modal expansion at higher squeezing amplitudes.
	
	\subsection{Two-mode Squeezed Scalar Beams}
	
	We construct optical analogs of two-mode squeezed states~\cite{MoralesRodriguez2024p1498, MoralesRodriguez2024} using coherent superpositions of LGBs,
	\begin{align}
		\Psi_{p,\ell,\zeta}(\rho,\theta,z) = \sum_{j=0}^{\infty} c_{j}(p, \vert \ell \vert, \alpha, \beta) \Psi_{j, \ell}(\rho,\theta,z),
	\end{align} 
	where the radial and azimuthal mode numbers, $p$ and $\ell$, and the complex squeezing parameter $\zeta = \alpha e^{i \beta}$, define the two-mode squeezing transformation. 
	
	The coefficients,
	\begin{align}
		\begin{aligned}
			c_{j}(p, \ell , \alpha, \beta) =&~ (-1)^{j} \sqrt{\binom{\vert \ell \vert + p - 1}{p} \binom{\vert \ell \vert + j - 1}{j}} \times \\
			&~ \times e^{i \beta (p - j)} \left( \mathrm{sech} \, \alpha \right)^{\vert \ell \vert + 1} \left( \mathrm{tanh} \, \alpha \right)^{p+j} \times \\
			&~ \times {}_{2}F_{1} \left( -j, -p; \vert \ell \vert + 1; -\mathrm{csch}^{2} \alpha \right),
		\end{aligned}
	\end{align}
	are the modal weights, expressed in terms of binomial coefficients $\binom{a}{b}$ and the Gauss hypergeometric function ${}_{2}F_{1}(a, b; c; d)$.
	
	For fixed squeezing parameter $\zeta$, the resulting squeezed LGBs form a new orthonormal basis. 
	As a result, the set of squeezed LGBs can represent any paraxial beam at that squeezing level, including optical analogs of two-mode Gaussian and non-Gaussian CV quantum states.
	
\begin{figure}
		\centering
		\includegraphics[scale = 1]{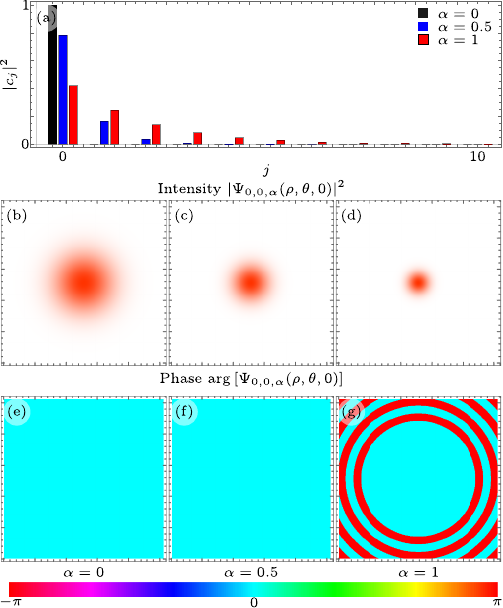}
		\caption{Two-mode squeezed LGB with $(p, \ell) = (0, 0)$. 
			(a) Modal weights $\vert c_{j}(p, \ell, \alpha, \beta)\vert^{2}$. (b)--(d) Intensity $\vert \Psi_{p,\ell,\zeta}(\rho,\theta,z) \vert^{2}$ at $z=0$, and (e)--(g) Phase $\mathrm{arg} \left[ \Psi_{p,\ell,\zeta}(\rho,\theta,z) \right]$ at $z=0$, for $\alpha \in \left\{ 0, 0.5, 1 \right\}$. 
			Beam parameters: $w_{0} = 1~\mathrm{mm}$, $\lambda = 632~\mathrm{nm}$, $\vert x \vert, \vert y \vert \leq 2 w_{0}$.}
		\label{fig:Fig4}
	\end{figure}

\begin{figure}
		\centering
		\includegraphics[scale = 1]{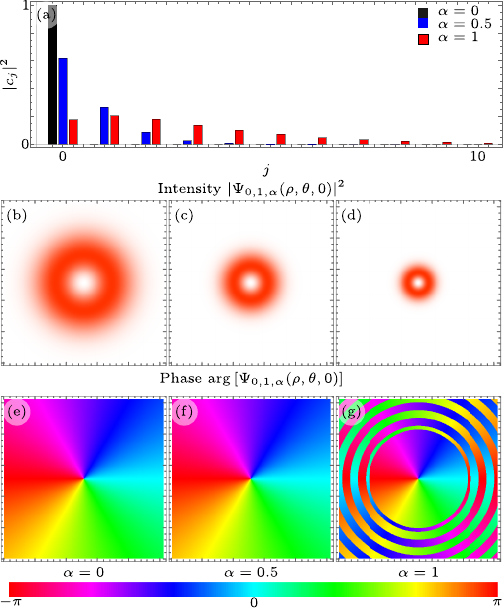}
		\caption{Same as Fig.~\ref{fig:Fig4}, but for a two-mode squeezed LGB with $(p, \ell) = (0, 1)$.}
		\label{fig:Fig5}
	\end{figure}

	Figure~\ref{fig:Fig4} shows a two-mode squeezed LGB with $(p, \ell) = (0, 0)$, corresponding to the two-mode squeezed vacuum state. 
	Figure~\ref{fig:Fig5} shows the case $(p, \ell) = (0, 1)$, representing the optical analog of a squeezed two-mode Fock state. 
	In both figures, panel (a) shows the squared modal weights $\vert c_j \vert^2$ for squeezing strengths $\alpha \in \{ 0, 0.5, 1 \}$; panels (b)--(d) show the intensity profiles; and panels (e)--(g) show the phase profiles at $z = 0$ for the same values of $\alpha$. 
	Panel (g) highlights the effects of truncating the modal expansion at higher squeezing amplitudes.
	
	\section{Discussion}
	\label{sec:S4}
	
	The Lie group-theoretic framework introduced in Sec.~\ref{sec:S3} provides a general description of squeezing in paraxial optics through transformations on structured beams. 
	This formulation enables the construction of optical analogs of Gaussian and non-Gaussian CV quantum states using coherent superpositions of paraxial modes.
	
	We use this framework to investigate core properties of squeezed beams. 
	In Sec.~\ref{PhaseRedist}, we  study their phase-space structure and establish explicit connections to CV quantum states. 
	We quantify the impact of squeezing on beam propagation in Sec.~\ref{DiffLimit}, where we derive analytical expressions for their diffraction-limited divergence  and minimum spot size. 
	In Sec.~\ref{ClassEntanglement}, we demonstrate that modal correlations and quadrature variances satisfy our classical inseparability criteria for two-mode squeezed beams.

	\subsection{Phase Space Representation}
	\label{PhaseRedist}
	
	\subsubsection{Squeezing of Fourier conjugate variables}
	
	In quantum mechanics, the two-dimensional Fourier transform operator,
	\begin{align}
		\hat{\mathcal{F}} = e^{-i \frac{\pi}{2} \sum_{j = x, y} \frac{1}{2} \left( \hat{p}_{j}^{2} - \hat{q}_{j}^{2} \right)},
	\end{align}
	acts as a $\pi/2$ rotation in each $(\hat{q}_j, \hat{p}_j)$ phase-space plane, transforming $\hat{q}_{j} \rightarrow \hat{p}_{j}$ and $\hat{p}_{j} \rightarrow -\hat{q}_{j}$. The same transformation occurs in optics, where the two-dimensional Fourier transform performs independent phase-space rotations between the real-space coordinates $(x, y)$ and spatial frequencies $(k_{x}, k_{y})$~\cite{Namias1980, McBride1987, Bailey1991}.
	
	Hermite-Gaussian and Laguerre-Gaussian modes are eigenfunctions of this operator,
	\begin{align}
		\begin{aligned}
			\hat{\mathcal{F}} \, \psi_{n_{x}, n_{y}}(q_{x}, q_{y}) &= (-i)^{n_{x} + n_{y} + 1} \, \psi_{n_{x}, n_{y}}(p_{x}, p_{y}), \\
			\hat{\mathcal{F}} \, \psi_{p, \ell}(q_{\rho}, q_{\theta}) &= (-i)^{2p + \vert \ell \vert + 1} \, \psi_{p, \ell}(p_{\rho}, p_{\theta}),
		\end{aligned}
	\end{align}
	preserving their functional structure and mapping the spatial coordinates to their Fourier-conjugate variables, up to a global phase determined by the total excitation number.
	
	Since squeezing acts as a hyperbolic transformation, Sec.~\ref{sec:S2}, compression of the spatial distribution along one direction results in expansion along its Fourier conjugate. 
	This behavior is consistent with the transformation of quadrature variances under single-mode squeezing, Eq.~(\ref{SqVariances}), where the variance in $q_{j}$ scales as $e^{\zeta_{j}}$ and in $p_{j}$ as $e^{-\zeta_{j}}$. 
	The product $\Delta q_{j} \Delta p_{j} \ge ( n + 1/2)$ remains bounded from below, in accordance with the uncertainty principle.
	
\begin{figure}
		\centering
		\includegraphics[scale = 1]{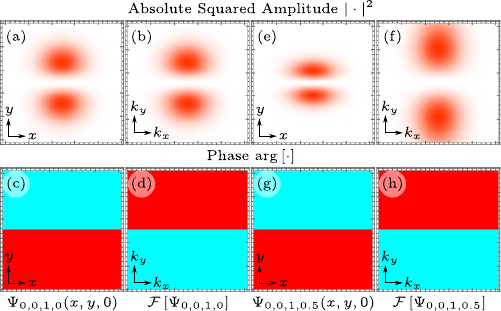}
		\caption{
			Real and Fourier space amplitude and phase of a single-mode squeezed HGB with $(n_{x}, n_{y}) = (0, 1)$ at $z = 0$. 
			(a)--(d): unsqueezed beam with $\alpha_{x} = \alpha_{y} = 0$, showing (a) intensity and (c) phase in real space, and (b) squared amplitude and (d) phase in Fourier space. 
			(e)--(h): squeezed beam with $\alpha_{x} = 0$ and $\alpha_{y} = 0.5$, showing (e) intensity and (f) phase in real space, and (g) squared amplitude and (h) phase in Fourier space.  
			Beam parameters: $w_{0} = 1~\mathrm{mm}$, $\lambda = 632~\mathrm{nm}$, with $\vert x \vert, \vert y \vert \le 2 w_{0}$ and $\vert k_{x} \vert, \vert k_{y} \vert \le 4/w_{0}$.
		}
		\label{fig:Fig6}
	\end{figure}

\begin{figure}
		\centering
		\includegraphics[scale = 1]{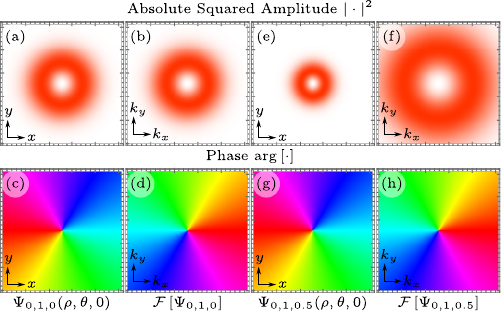}
		\caption{
			Same as Fig.~\ref{fig:Fig6}, for a two-mode squeezed LGB with $(p, \ell) = (0, 1)$ and squeezing amplitude $\alpha = 0.5$.
		}
		\label{fig:Fig7}
	\end{figure}

	Figure~\ref{fig:Fig6} shows a single-mode squeezed Hermite-Gaussian beam with $(n_{x}, n_{y}) = (0, 1)$, and Fig.~\ref{fig:Fig7} shows a two-mode squeezed Laguerre-Gaussian beam with $(p, \ell) = (0, 1)$. 
	In both cases, panels (a) and (c) display the intensity and phase in real space $(x, y)$ at $z = 0$ for the unsqueezed beam with $\alpha_{x} = \alpha_{y} = \alpha = 0$, and panels (b) and (d) show the corresponding squared amplitude and phase in Fourier space. 
	Panels (e)--(h) present the same information for the squeezed cases, with $\alpha_{x} = 0$ and $\alpha_{y} = 0.5$ in Fig.~\ref{fig:Fig6}, and $\alpha = 0.5$ in Fig.~\ref{fig:Fig7}.
	In both cases, the intensity  of the field compresses in real space, while the squared amplitude expands in Fourier space, analogous to squeezing and antisqueezing of conjugate variables. 
    This behavior is consistent with hyperbolic phase-space redistribution and  illustrates the trade-off between spatial and spectral localization imposed by the uncertainty principle.

	\subsubsection{Optical Wigner distribution}
	
	Fourier-conjugate variables reveal structural analogies between classical and quantum squeezing, namely the compression and expansion of variances of conjugate variables. 
	However, this description does not fully capture the correspondence with quantum states; for example, it omits information about phase correlations and spatial coherence.
	The optical Wigner distribution function encodes both spatial and spectral structure of the beam in phase space, offering a more informative representation than real- or Fourier-space scalar fields on their own~\cite{Bastiaans1985, Dragoman1997}.
	
	For a two-dimensional scalar beam, we define the optical Wigner distribution,
	\begin{align}
		\begin{aligned}
			W(x, y, k_{x}, k_{y}) =&~ \frac{1}{(2\pi)^2} \iint_{-\infty}^{\infty} \mathrm{d}u\, \mathrm{d}v \, e^{-i (k_{x} u + k_{y} v)} \times \\
			&~ \times \Psi^*\left(x - \frac{u}{2}, y - \frac{v}{2} \right) \Psi\left(x + \frac{u}{2}, y + \frac{v}{2} \right),
		\end{aligned}
	\end{align}
	which is a real-valued function over real space and spatial frequency that captures spatial coherence and phase correlations~\cite{Alonso2011}. 
	It provides a phase-space framework to investigate correlations in classical analogs of CV quantum states.
	
	For squeezed HGBs and LGBs, the optical Wigner distribution function undergoes hyperbolic deformations along the squeezed and anti-squeezed quadratures, while preserving total phase-space area. 
    This redistribution contains information about how squeezing modulates the spatial–spectral structure of the beam.
	
	For an arbitrary scalar field in the HGBs basis, 
	\begin{align}
		\Psi(x,y,z) = \sum_{m,n=0}^{\infty} c_{m,n} \Psi_{m,n}(x,y,z),
	\end{align}
	the Wigner distribution is a superposition,
	\begin{align}
		W(x,k_{x},y,k_{y}) = \sum_{m,n=0}^{\infty} \sum_{l,i=0}^{\infty} c_{m,n}^{\ast} c_{l,i} \, W_{m,n;l,i}(x,k_{x},y,k_{y}),
	\end{align}
	of mode-resolved contributions,
	\begin{align}
		\begin{aligned}
			&~ W_{m,n;l,i}(x,k_{x},y,k_{y}) = \frac{e^{-i(m+n-l-i)\varphi(z)}}{\pi^{2} \sqrt{m! \, n! \, l! \, i!}} \, e^{- \frac{2}{w^{2}(z)} \left(x^{2} + y^{2}\right)} \times \\
			&~ \times e^{- \frac{1}{4} \left[ K^{2}(x) + K^{2}(y) \right]} \, H_{m,l}\left[u(x), v(x)\right] \, H_{n,i}\left[u(y), v(y)\right].
		\end{aligned}
	\end{align}
	Here, we used the bivariate Hermite polynomials,
	\begin{align}
		H_{m,n}(x,y) = \left[ \frac{d^{m}}{ds^{m}} \frac{d^{n}}{dt^{n}} e^{sx + ty - st} \right]_{s=t=0},
	\end{align}
	and the auxiliary functions,
	\begin{align}
		\begin{aligned}
			K(\zeta) =&~ k_{\zeta} - \frac{2k}{R(z)} \zeta, \\
			u(\zeta) =&~ \frac{1}{\sqrt{2}} \left[ \frac{2 \sqrt{2} \zeta}{w(z)} + i K(\zeta) \right], \\
			v(\zeta) =&~ \frac{1}{\sqrt{2}} \left[ \frac{2 \sqrt{2} \zeta}{w(z)} - i K(\zeta) \right],
		\end{aligned}
	\end{align}
	for $\zeta = x, y$, that encode the beam curvature and shearing in phase space.
	
    The optical Wigner distribution function provides a versatile and useful phase-space framework to analyze classical analogs of CV Gaussian and non-Gaussian states. 
	For example, the HGB with modal numbers $(n_{x}, n_{y})$ is the classical analog of the Fock state $\vert n_{x}, n_{y} \rangle$ of the two-dimensional quantum harmonic oscillator, which is non-Gaussian for $n_{x}, n_{y} > 0$. 
	Both the classical and quantum Wigner distributions exhibit negative regions, although the physical interpretation of their negativity differs. 
	In the classical case, Wigner negativity results from phase-space interference produced by the nodal pattern of the field. 	
    In the quantum case,  Wigner negativity indicates nonclassicality and reflects the non-Gaussian character of the state. 
	
\begin{figure}
		\centering
		\includegraphics[scale = 1]{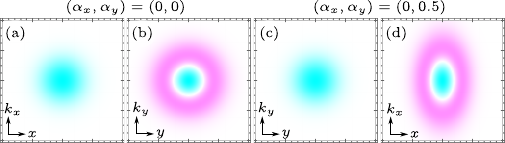}
		\caption{
			Optical Wigner distribution in the planes (a) $(x, k_{x})$ for $y = k_{y} = 0$ and (b) $(y, k_{y})$ for $x = k_{x} = 0$ for an HGB with $(n_{x}, n_{y}) = (0, 1)$, evaluated at $z = 0$. 
			Left panels show the unsqueezed case; right panels show squeezing with $\alpha_{y} = 0.5$. 
			Beam parameters: $w_{0} = 1~\mathrm{mm}$, $\lambda = 632~\mathrm{nm}$, with $\vert x \vert, \vert y \vert \le 2 w_{0}$ and $\vert k_{x} \vert, \vert k_{y} \vert \le 4/w_{0}$.
		}
		\label{fig:Fig8}
	\end{figure}

	Figure~\ref{fig:Fig8} shows the Wigner distribution of a HGB with modal numbers $(n_{x}, n_{y}) = (0, 1)$ projected onto different phase-space planes, with and without squeezing along $y$. 
	Panels (a) and (c) show the $(x, k_{x})$ plane, evaluated at $y = k_{y} = 0$, while panels (b) and (d) show the $(y, k_{y})$ plane, evaluated at $x = k_{x} = 0$. 
	Panels (a) and (b) correspond to the unsqueezed case, and panels (c) and (d) show the squeezed case with $\alpha_{x} = 0$ and $\alpha_{y} = 0.5$, i.e. squeezed along $y$. 
	The deformation in the $(y, k_{y})$ plane illustrates the redistribution of phase-space components induced by squeezing.
	In both cases, negativity arises from interference, which in classical optics is determined by the mode geometry, while in the quantum analog indicates non-Gaussianity.
	
	\subsection{Diffraction Limit}
	\label{DiffLimit}

We examine whether squeezed optical beams can improve imaging resolution by analyzing their propagation through finite apertures.  
The diffraction limit constrains spatial resolution to the smallest achievable spot size for a given wavelength and numerical aperture.

In paraxial propagation through homogeneous media, a circular aperture imposes a radial cut-off in spatial frequency space,  
\begin{align}
    k_{\rho, \max}^{\mathrm{(A)}} =&~ \frac{2 \pi}{\lambda_{0}} \mathrm{NA}^{\mathrm{(A)}},
\end{align}
where $\lambda_{0}$ is the free-space wavelength.  
We express the numerical aperture as a function of the refractive index $n$, aperture radius $a$, and propagation distance $z$,  
\begin{align}
    \mathrm{NA}^{\mathrm{(A)}} =&~ n \sin \theta = \frac{n a}{\sqrt{a^{2} + z^{2}}} \approx \frac{n a}{z},
\end{align}
where the far-field approximation holds for $a \ll z$.  
Although the NA is defined using the far-field angle $\theta$, the cut-off applies uniformly at all propagation distances.  
We choose this spatial frequency cut-off to characterize the diffraction limit due to this propagation invariance.

A fundamental Gaussian beam has a spatial frequency cut-off,  
\begin{align}
    k_{\rho, \max}^{\mathrm{(G)}} =  \frac{2}{ w_{0}} \frac{1}{\sqrt{1+ \left(  \frac{\lambda_{0}}{n \pi  w_{0}} \right)^{2}}} \approx \frac{2}{w_{0}},
\end{align}
which depends only on the beam waist $w_{0}$ in the small-angle limit $\theta \ll 1$.  
The corresponding numerical aperture,  
\begin{align}
    \mathrm{NA}^{\mathrm{(G)}} =  \frac{\lambda_{0}}{\pi w_{0}} \frac{1}{\sqrt{1+ \left(  \frac{\lambda_{0}}{n \pi  w_{0}} \right)^{2}}} \approx \frac{\lambda_{0}}{\pi w_{0}},
\end{align}
provides the NA that a circular aperture must match to reproduce the same spatial frequency cut-off as the Gaussian beam.

Our key question is whether squeezing can reduce the beam waist below the Gaussian beam waist $w_{0}$ for a fixed NA provided by a circular aperture. To analyze this problem, we work in the LGB basis and examine the effect of squeezing on the fundamental mode.  
For a two-mode squeezed LGB with $(p, \ell) = (0, 0)$, the field remains Gaussian,
\begin{align}
	\Psi_{0,0,\zeta} (\rho,\theta,z) = \frac{1}{\sqrt{\pi}} \frac{\sqrt{2}}{w(z; \alpha, \beta)}  
	e^{ -\frac{i k \rho^{2}}{2 R(z; \alpha, \beta)} } 
	e^{ -i \varphi(z; \alpha, \beta)} 
	e^{ - \frac{\rho^{2}}{w^{2}(z; \alpha, \beta)} },
\end{align}
and we obtain analytic expressions for the modified waist, curvature, and Gouy phase,
\begin{align}
	\begin{aligned}
		w(z; \alpha, \beta) =&~ \left\vert 1 - e^{-i 2 \varphi(z)} e^{i \beta} \mathrm{coth} \, \alpha \right\vert \mathrm{sinh} \, \alpha \, w(z), \\
	{ R(z; \alpha, \beta)} =&~{   \frac{ \left\vert 1 - e^{-i 2 \varphi(z)} e^{i \beta} \mathrm{coth} \, \alpha \right\vert^{2} w^{2}(z)  R(z)}{\left\vert 1 - e^{-i 2 \varphi(z)} e^{i \beta} \mathrm{coth} \, \alpha \right\vert^{2} w^{2}(z) + 4 \sin \left[ 2 \varphi(z) - \beta \right] \mathrm{coth} \, \alpha },} \\
		\varphi(z; \alpha, \beta) =&~ \varphi(z) + \mathrm{arctan} \left( \frac{\sin \left[ 2 \varphi(z) - \beta \right] \mathrm{coth} \, \alpha}{\tanh \alpha - \cos \left[ 2 \varphi(z) - \beta \right]} \right),
	\end{aligned}
\end{align}
with minimum waist at $z = 0$ when $\beta = 0$.  
We approximate the Rayleigh length using the squeezed waist,
\begin{align}
	z_{R}(\alpha, \beta) \approx \frac{\pi}{\lambda} w^{2}(0; \alpha, \beta),
\end{align}
to model squeezing effects under standard Gaussian beam propagation.
	
\begin{figure}
		\centering
		\includegraphics[scale = 1]{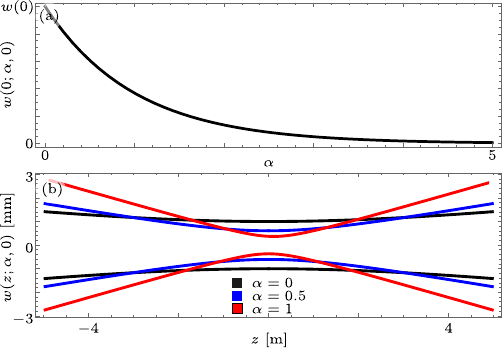}
		\caption{Beam waist for squeezed LGB with $(p, \ell) = (0, 0)$.
			(a) minimum beam waist $w(z = 0; \alpha, \beta = 0)$ as a function of squeezing amplitude $\alpha$,
			(b) $w(z; \alpha, \beta = 0)$ for propagating LGB for $\alpha = 0, 0.5, 1$. 
			Beam parameters: $w_{0} = 1~\mathrm{mm}$, $\lambda = 632~\mathrm{nm}$.}
		\label{fig:Fig9}
	\end{figure}

Figure~\ref{fig:Fig9}(a) shows the minimum waist $w(z = 0; \alpha, \beta = 0)$ of a two-mode squeezed LGB with $(p, \ell) = (0, 0)$ as a function of the squeezing amplitude $\alpha$.  
Figure~\ref{fig:Fig9}(b) shows the corresponding propagation of $w(z; \alpha, \beta = 0)$ for $\alpha \in \{0, 0.5, 1\}$.  
In both panels, we present our ideal analytic results without any circular aperture stop.  
Squeezing reduces the waist at $z = 0$ and increases the far-field divergence relative to the unsqueezed beam with $\alpha = 0$.

We numerically investigated the propagation of high-order squeezed beams for varying squeezing strengths.  
Under propagation, the waist,
\begin{align}
	w_{M}(z; \alpha, \beta) = M^{2} w(z; \alpha, \beta),
\end{align}
is directly proportional to the beam quality factor,
\begin{align}
	M^{2} = n_{x} + n_{y} + 1 = 2p + \vert\ell\vert + 1,
\end{align}
which is determined by the modal numbers as in standard Gaussian optics.  
This scaling implies that a system with finite NA supports only a limited number of modes.  
Since squeezing draws from higher-order modes that diverge more rapidly, a circular aperture bounds the achievable squeezing and the minimum waist at focus.

\begin{figure}
		\centering
		\includegraphics[scale = 1]{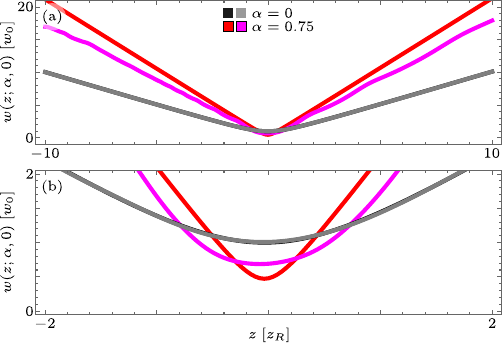}
		\caption{ Beam waist for squeezed LGB with $(p, \ell) = (0, 0)$.
			(a) Beam waist $w(z; \alpha, \beta = 0)$ for $\alpha = 0$ and $\alpha = 0.75$, along with their apodized profiles truncated by a circular aperture of radius $2 w(-10 z_{R})$. 
			(b) Zoomed view of the region $z \in [-2 z_{R}, 2 z_{R}]$.
		}
		\label{fig:Fig10}
	\end{figure}

Figure~\ref{fig:Fig10} shows the effect of a finite NA on the focusing behavior of squeezed beams.  
Panel (a) plots the ideal waist evolution for LGBs with modal numbers $p = \ell = 0$, showing the unsqueezed (black) and squeezed (red) beams without any aperture.  
The gray and magenta curves show the corresponding apodized beams truncated by a circular aperture of radius $2 w(-10 z_{R})$.  
This radius is sufficient to match the fundamental Gaussian beam.  
Panel (b) zooms into the focal region near $z = 0$, where the truncation effect on the squeezed beam becomes evident.
The finite numerical aperture limits the number of accessible modes, setting a bound on the achievable squeezing and the minimum waist at the plane $z = 0$.  
As can be seen from Figure~\ref{fig:Fig10}, an aperture with a spatial frequency cut-off comparable to that of the fundamental Gaussian mode can still accommodate moderate squeezing with minimal distortion.

	\subsection{Quantum-inspired inseparability for optical beams: Continuous-variable classical entanglement}
	\label{ClassEntanglement}
	
	Classical entanglement~\cite{Barut1984} emerges in structured optical fields~\cite{Forbes2019}, where correlations between spatial modes mirror criteria for DV quantum inseparability~\cite{Kaszlikowski2008}. 
    In the CV regime, two-mode squeezing induces quadrature correlations that can produce entanglement. 
    The Duan–Simon criterion ~\cite{Duan2000} provides a necessary and sufficient condition of separability  of Gaussian states. Using the EPR-type operators,
	\begin{align}
		\hat{u} = a \hat{q}_x + \frac{1}{a} \hat{q}_y, \qquad 
		\hat{v} = a \hat{p}_x - \frac{1}{a} \hat{p}_y,
	\end{align}
	where $a \in \mathbb{R}$, this criterion states that for separable states the sum of variances of these operators satisfy the inequality,
	\begin{align}
		\langle (\Delta \hat{u})^2 \rangle + \langle (\Delta \hat{v})^2 \rangle \ge a^2 + \frac{1}{a^2}.
	\end{align}
	A violation of this bound indicates two-mode entanglement, that is, inseparability between the modes.
	For non-Gaussian states, such as squeezed number states, the criterion remains sufficient but not necessary. 
	As a result, some non-Gaussian entangled states satisfy the inequality, and detecting their inseparability requires relying on other phase-space signatures or alternative witnesses.
	
	A common simplification of the Duan–Simon criterion sets $a = 1$,
	\begin{align}
		\hat{u} = \hat{q}_x + \hat{q}_y, \qquad
		\hat{v} = \hat{p}_x - \hat{p}_y.
		\label{eq:epr_operators}
	\end{align}
	These combinations correspond to diagonal coordinate and anti-diagonal momentum observables. 
    Under two-mode squeezing,
	\begin{align}
		e^{i \zeta_{2} \hat{K}_{2}} \hat{u} \, e^{-i \zeta_{2} \hat{K}_{2}} &= \hat{q}_x e^{-\zeta_{2}/2} + \hat{q}_y e^{\zeta_{2}/2}, \\
		e^{i \zeta_{2} \hat{K}_{2}} \hat{v} \, e^{-i \zeta_{2} \hat{K}_{2}} &= \hat{p}_x e^{-\zeta_{2}/2} - \hat{p}_y e^{\zeta_{2}/2},
	\end{align}
	the modes rescale unevenly, distorting the original symmetric and antisymmetric combinations.
	The corresponding total variance of these EPR-type operators,
	\begin{align}
		\begin{aligned}
			&\langle \Delta^{2} ( \hat{q}_x + \hat{q}_y ) \rangle + \langle \Delta^{2}( \hat{p}_x - \hat{p}_y )\rangle  = \\ & e^{-\zeta_2} \left[ \langle \hat{q}_x^2 \rangle + \langle \hat{q}_y^2 \rangle  + \langle \hat{p}_x^2 \rangle + \langle \hat{p}_y^2 \rangle   - 2 \langle \hat{q}_x \hat{q}_y \rangle - 2 \langle \hat{p}_x \hat{p}_y \rangle \right],
		\end{aligned}
		\label{eq:variance_transformed}
	\end{align}
	yields the left-hand-side of the simplified Duan–Simon inseparability condition for CV Gaussian states,
	\begin{align}
		\langle \Delta^{2} ( \hat{q}_x + \hat{q}_y ) \rangle + \langle \Delta^{2}( \hat{p}_x - \hat{p}_y )\rangle  \ge 2.
		\label{eq:duan_simon_classical}
	\end{align}
	
    \subsubsection{Optical analog of inseparability criterion}
	
	We construct a classical optical analog of the simplified Duan–Simon criterion for CV Gaussian states using two-mode squeezed Hermite-Gaussian beams. The variances for the spatial coordinates $(\hat{q}_{1}, \hat{q}_{2}) \rightarrow (x, y)$ and their conjugate momenta $(\hat{p}_{1}, \hat{p}_{2}) \rightarrow (-i \partial_{x}, -i \partial_{y})$,
	\begin{align}
		\langle \Delta^{2} (x + y) \rangle &= \left[ \langle x^{2} \rangle + \langle (-i \partial_{x})^{2} \rangle \right] e^{-\zeta}, \\
		\langle \Delta^{2} (-i \partial_{x} + i \partial_{y}) \rangle &= \left[ \langle y^{2} \rangle + \langle (-i \partial_{y})^{2} \rangle \right] e^{-\zeta},
		\label{eq:classical_variances}
	\end{align}
	follow from second moments evaluated on the spatial profile,
	\begin{align}
		\langle o^{n} \rangle = \iint_{-\infty}^{\infty} \mathrm{d}x \, \mathrm{d}y \, \Psi_{n_{x},n_{y}}^{*}(x,y,z) \, o^{n} \, \Psi_{n_{x},n_{y}}(x,y,z),
		\label{eq:moments}
	\end{align}
	with $o = x + y$ or $o = -i \partial_{x} + i \partial_{y}$.
    The first moments vanish,
	\begin{align}
		\begin{aligned}
			\langle x + y \rangle =&~ 0, \\
			\langle -i \partial_{x} + i \partial_{y} \rangle =&~ 0,
		\end{aligned}
		\label{eq:first_moments}
	\end{align}
	as expected from HGBs optical modes. The second moments,
	\begin{align}
		\begin{aligned}
			\langle (x + y)^{2} \rangle =&~ (n_{x} + n_{y} + 1) e^{-\zeta}, \\
			\langle (-i \partial_{x} + i \partial_{y})^{2} \rangle =&~ (n_{x} + n_{y} + 1) e^{-\zeta},
		\end{aligned}
		\label{eq:second_moments}
	\end{align}
	lead to the left-hand side of our classical analog of the simplified Duan–Simon inequality,
	\begin{align}
		\langle \Delta^{2} (x + y) \rangle + \langle \Delta^{2} (-i \partial_{x} + i \partial_{y}) \rangle \ge (n_{x} + n_{y} + 1) e^{-\zeta},
		\label{eq:classical_duan_simon}
	\end{align}
	This results in the inseparability criterion
	\begin{align}
		\langle \Delta \hat{u}^{2} \rangle + \langle \Delta \hat{v}^{2} \rangle = e^{-\zeta} (n_{x} + n_{y} + 1) \ge 1,
		\label{eq:classical_duan}
	\end{align}
	for two-mode squeezed Hermite-Gaussian beams.
	Our inequality in Eq. (\ref{eq:classical_duan_simon}) generalizes the Duan–Simon criterion and recovers the inseparability condition for the classical analog of two-mode squeezed vacuum states, i.e. the squeezed HGB $\Psi_{n_{x},n_{y},\zeta}(x,y,z)$ with $(n_{x}, n_{y}) = (0,0)$. 
	For this beam, the condition is necessary and sufficient, for which any $\zeta > 0$ implies a violation of Eq.~\eqref{eq:classical_duan_simon}, confirming continuous-variable classical entanglement.
	On the other hand, for the classical analogs of squeezed number states, $\Psi_{n_{x},n_{y},\zeta}(x,y,z)$ with $(n_{x}, n_{y}) \ne (0,0)$, the criterion remains necessary but not sufficient. Therefore, for these non-Gaussian beams satisfying the inequality does not rule out the possibility of entanglement.

	\subsubsection{Marginal optical Wigner distribution functions}
	
	The optical Wigner distribution function provides an alternative witness of correlations in two-mode squeezed optical scalar beams. 
	We consider the four marginal Wigner distributions,
	\begin{align}
		W(x, k_{x}) &= \iint_{-\infty}^{\infty} \mathrm{d}y \, \mathrm{d}k_{y} ~ W(x, k_{x}, y, k_{y}), \nonumber \\
		W(y, k_{y}) &= \iint_{-\infty}^{\infty} \mathrm{d}x \, \mathrm{d}k_{x} ~ W(x, k_{x}, y, k_{y}), \nonumber \\
		W(x, y)     &= \iint_{-\infty}^{\infty} \mathrm{d}k_{x} \, \mathrm{d}k_{y} ~ W(x, k_{x}, y, k_{y}), \nonumber \\
		W(k_{x}, k_{y}) &= \iint_{-\infty}^{\infty} \mathrm{d}x \, \mathrm{d}y ~ W(x, k_{x}, y, k_{y}),
		\label{eq:marginalWigner}
	\end{align}
	that encode partial phase-space information and reveal correlations between specific conjugate or transverse degrees of freedom~\cite{Dragoman1997, Bastiaans1985, Alonso2011}.
	The hybrid distributions $W(x, k_{x})$ and $W(y, k_{y})$ capture the interplay between real– and Fourier-space variables, while $W(x, y)$ and $W(k_{x}, k_{y})$ capture correlations within purely spatial or spectral domains.

\begin{figure}
    \centering
    \includegraphics[scale = 1]{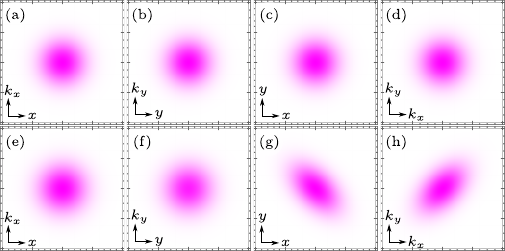}
    \caption{Marginal Wigner distributions (a) $W(x,k_{x})$, (b) $W(y,k_{y})$, (c) $W(x,y)$, and (d) $W(k_{x},k_{y})$ for an HGB with $n_{x}=n_{y}=0$ at $z=0$: without squeezing (first row); and with two-mode squeezing with squeezing amplitude $\alpha=0.5$ (second row). The Gaussian beam parameters are $w_{0}=1~\mathrm{mm}$ and $\lambda = 632~\mathrm{nm}$, with plot ranges $ \vert x \vert, \vert y \vert \le 2 w_{0} $ and $\vert k_{x} \vert, \vert k_{y}\vert \leq 4 / w_{0} $.}
    \label{fig:Fig11}
    \end{figure}

	We illustrate the structure of the marginal Wigner distributions using the unsqueezed fundamental HGB, $(n_{x}, n_{y}) = (0, 0)$, at $z = 0$, 
	\begin{align}
		\Psi_{0,0}(x, y, 0) = \frac{\sqrt{2}}{w_{0}} e^{- \frac{x^{2} + y^{2}}{w_{0}^{2}}},
		\label{eq:fundamentalHGB}
	\end{align}
	with optical Wigner distribution function,
	\begin{align}
		W(x, k_{x}, y, k_{y}) = \frac{1}{\pi} e^{ - \frac{2}{w_{0}^{2}} (x^{2} + y^{2}) } e^{ - \frac{w_{0}^{2}}{2} (k_{x}^{2} + k_{y}^{2}) },
		\label{eq:WignerFundamental}
	\end{align}
	and marginal distributions,
	\begin{align}
		W(x, k_{x}) &= e^{- \frac{2}{w_{0}^{2}} x^{2}} \, e^{- \frac{w_{0}^{2}}{2} k_{x}^{2}}, \nonumber \\
		W(y, k_{y}) &= e^{- \frac{2}{w_{0}^{2}} y^{2}} \, e^{- \frac{w_{0}^{2}}{2} k_{y}^{2}}, \nonumber \\
		W(x, y)     &= \frac{2}{w_{0}^{2}} e^{- \frac{2}{w_{0}^{2}} (x^{2} + y^{2})}, \nonumber \\
		W(k_{x}, k_{y}) &= \frac{w_{0}^{2}}{2} e^{- \frac{w_{0}^{2}}{2} (k_{x}^{2} + k_{y}^{2})}.
		\label{eq:marginalsNoSqueeze}
	\end{align}
	Figure~\ref{fig:Fig11}(a)--(d) shows these marginal distributions, which are all separable in products of Gaussians in their respective variables.
	
	We obtain the analytic expression for the two-mode squeezed fundamental HGB, $(n_{x}, n_{y}) = (0, 0)$, at $z = 0$,
	\begin{align}
		\Psi_{0,0,\alpha}(x, y, 0) = \frac{\sqrt{2}}{w_{0}} 
		e^{- \frac{1}{w_{0}^{2}} \left[ (x^{2} + y^{2}) \cosh \alpha + 2 x y \sinh \alpha \right]},
		\label{eq:squeezedPsi00}
	\end{align}
	with corresponding optical Wigner distribution function,
	\begin{align}
		\begin{aligned}
			W(x, k_{x}, y, k_{y}) = \frac{1}{\pi} 
			&\, e^{- \frac{2}{w_{0}^{2}} \left[ (x^{2} + y^{2}) \cosh \alpha + 2 x y \sinh \alpha \right]} \\
			&\, \times e^{- \frac{w_{0}^{2}}{2} \left[ (k_{x}^{2} + k_{y}^{2}) \cosh \alpha - 2 k_{x} k_{y} \sinh \alpha \right]},
		\end{aligned}
		\label{eq:fullWigner}
	\end{align}
	leading to closed-form expressions for the marginal distributions,
	\begin{align}
		W(x, k_{x}) &= \mathrm{sech} \, \alpha \, 
		e^{- \frac{2}{w_{0}^{2}} \, \mathrm{sech} \, \alpha \, x^{2}} 
		e^{- \frac{w_{0}^{2}}{2} \, \mathrm{sech} \, \alpha \, k_{x}^{2}}, \nonumber \\
		W(y, k_{y}) &= \mathrm{sech} \, \alpha \, 
		e^{- \frac{2}{w_{0}^{2}} \, \mathrm{sech} \, \alpha \, y^{2}} 
		e^{- \frac{w_{0}^{2}}{2} \, \mathrm{sech} \, \alpha \, k_{y}^{2}}, \nonumber \\
		W(x, y) &= \frac{2}{w_{0}^{2}} 
		e^{- \frac{2}{w_{0}^{2}} \left[ (x^{2} + y^{2}) \cosh \alpha + 2 x y \sinh \alpha \right]}, \nonumber \\
		W(k_{x}, k_{y}) &= \frac{w_{0}^{2}}{2} 
		e^{- \frac{w_{0}^{2}}{2} \left[ (k_{x}^{2} + k_{y}^{2}) \cosh \alpha - 2 k_{x} k_{y} \sinh \alpha \right]},
		\label{eq:marginalsSqueezed}
	\end{align}
	
    These marginal distributions capture quadrature correlations, which appear as correlated distortions of the separable Gaussian profiles in the $(x,y)$ and $(k_x,k_y)$ planes, as shown in Fig.~\ref{fig:Fig11}(e)--(h).

    These phase-space correlations constitute a signature of classical inseparability, establishing a direct analog of two-mode squeezed quantum states. 
    Our results provide a structured-light platform for exploring classical entanglement in the continuous-variable regime.
	
	\section{Conclusion}
	\label{Conclusions}
	
	We presented a comprehensive algebraic framework based on the $\mathrm{su}(1,1)$ Lie algebra to describe single- and two-mode squeezing in paraxial optics.  
	We constructed squeezed scalar beams as optical analogs of quantum squeezed number states and showed that our framework applies to arbitrary single- or two-mode representations.  
	The resulting beams form a complete and orthonormal basis for structured optical fields.
	
	Our analysis shows that optical squeezing redistributes phase space, modifying both spatial and spectral structure, and altering beam divergence and diffraction properties.  
	We derived analytic expressions for the scaled waist, curvature, and Gouy phase of two-mode squeezed Laguerre-Gaussian beams to quantify these effects.  
	These expressions reveal that the degree of spatial compression through squeezing remains bounded by the numerical aperture of the optical system, which limits the number of accessible higher-order modes and degree of achievable squeezing.  
	Digital holography and other mode-shaping techniques remain constrained by this bound.  
	Within these limits, coherent superpositions of orthonormal modes support spatially compressed beams compatible with, and limited by, the system's numerical aperture.
	
	We established a classical analog of the Duan–Simon inseparability criterion, which is  a necessary and sufficient  condition for the separability of continuous-variable Gaussian states.  
	For non-Gaussian optical analogs, marginals of the optical Wigner distribution reveal phase-space features, such as negativity, that serve as witnesses of classical inseparability and entanglement.  
	These results show that squeezed scalar beams exhibit modal correlations that reproduce key signatures of  continuous-variable quantum entanglement in a classical setting. 
    Our algebraic framework offers a path toward understanding and developing continuous-variable quantum-inspired applications in optical imaging, metrology, and communication.

    \section*{Funding}
    This work was partially funded by the National Institute of Health (NIH) Grant No. 1R01GM140284-01, the National Science Foundation (NSF) Grant No. PHY-2210447, and the Department of Energy (DOE) Contract No. CW42943.
	
	\section*{Acknowledgments}	
	B.~M.~R.~L. acknowledges support and hospitality as an affiliate visiting colleague at the Department of Physics and Astronomy, University of New Mexico and fruitful discussion with Dilia Aguirre-Olivas, Gabriel Mellado-Villase\~nor, and Benjamin de Jesus Perez Garcia.
	
	\section*{Disclosures} 
	The authors declare no conflicts of interest.
	
	\section*{Data availability} 
	Data underlying the results presented in this paper may be obtained from the authors upon reasonable request.
	

%

\end{document}